\documentclass{article}
\usepackage[english]{babel}
\usepackage[utf8]{inputenc}
\usepackage{multirow}
\usepackage{johd}
\usepackage{amsmath}
\usepackage{amssymb}
\usepackage{mathrsfs}
\usepackage{dsfont}
\usepackage{bm}
\usepackage{xcolor}
\usepackage[normalem]{ulem}
\usepackage{caption}
\DeclareUnicodeCharacter{0301}{\'{e}}

\usepackage{blindtext}

\title{Experimental optical simulator of \\ reconfigurable and complex quantum environment}

\author{P. Renault$^{1,*}$, J. Nokkala$^{2}$, G. Roeland $^{1}$, N. Y. Joly $^{5,6,7}$, \\ R. Zambrini$^{3}$, S. Maniscalco$^{4}$, J. Piilo$^{2}$  N. Treps$^{1}$, V. Parigi$^{1}$ \\
        \small $^{1}$Laboratoire Kastler Brossel, Sorbonne Université, CNRS, ENS-Université PSL, Collège de France, 75005 Paris \\
        \small $^{2}$Department of Physics and Astronomy, University of Turku,\\
        \small FI-20014 Turun yliopisto, Finland\\
        \small $^{3}$Instituto de F́ısica Interdisciplinar y Sistemas Complejos IFISC (CSIC-UIB), Campus Universitat Illes Balears, \\
        \small  E-07122 Palma de Mallorca, Spain\\
        \small $^{4}$QTF Centre of Excellence, Department of Physics,
University of Helsinki, P.O. Box 43, FI-00014 Helsinki, Finland,\\ 
         \small $^{5}$ Department of Physics, Friedrich-Alexander-Universität, Staudtstraße 2, 91058 Erlangen, Germany\\
         \small $^{6}$Max Planck Institute for the Science of Light, Staudtstraße 2, 91058 Erlangen Germany\\
         \small $^{7}$Interdisciplinary Centre for Nanostructured Films, Cauerstr.3, 91058 Erlangen, Germany
         \\
        \small $^{*}$ current address: Department of Physics, University of Virginia, 382 McCormick Rd, Charlottesville, 22904-4714, VA, USA
}

\date{} 

\begin{document}

\maketitle
\begin{abstract}
\noindent No quantum system can be considered totally isolated from its environment. In most cases the interaction between the system of interest and the external degrees of freedom deeply changes its dynamics, as described by open quantum system theory. Nevertheless  engineered environment can be turned into beneficial effects for some quantum information tasks.  Here we demonstrate an optical simulator of a quantum system coupled to an arbitrary and  reconfigurable environment built as a complex network of quantum interacting systems. We experimentally retrieve typical features of open quantum system dynamics like the spectral density and quantum non-Markovianity, by exploiting squeezing and entanglement correlation of a continuous variables optical platform. This opens the way to the experimental tests of open quantum systems in reconfigurable environments that are relevant in, among others,  quantum information, quantum thermodynamics, quantum transport and quantum synchronization.  
\end{abstract}
\section{Introduction}

Quantum information technologies are nowadays getting to the regime where noisy intermediate scale quantum systems (NISQ) \cite{Preskill2018} show quantum advantage when compared with classical equivalents \cite{Zhong2020,Arute2019,Wu2021, Madsen2022}.\\
Nevertheless, whatever the platform considered, both decoherence and losses are still obstacles that  can be  mitigated but not completely avoided \cite{Berdou2022,Lescanne2020}, so any platform should be considered an open system.  The theory of open quantum systems largely explored the role of the environment and showed that, when opportunely engineered, it can be promoted to be an ally of the open system \cite{Biercuk2009, Verstraete2009, Barreiro2010, Albert2016,Koch2016,Liu2016,Liu2022}.  \\
Beyond quantum information platforms, the study of open quantum systems and a structured environment is essential for understanding biological systems \cite{Mattioni21,Caycedo2022, Nussuler22}, boosting quantum thermal machines \cite{Manzano19,Kloc21} and machine learning protocol \cite{Sannia2022},  achieving collective phenomena such as dissipative phase transitions \cite{Minganti2018} or synchronization in the quantum realm \cite{Manzano13,Cabot19} and explaining the emergence of the classical world from quantum constituents \cite{Galve16,Le19,Foti19}.  It is thus crucial to have  experimental platforms to put to the test the variety of open quantum systems for different purposes. In this work we demonstrate an optical simulator of arbitrary quantum environments interacting with an open system. \\
The fine-grained structure of the environment can be described in many  cases as a network of quantum harmonic oscillators.   We experimentally reproduce the dynamics of the open system coupled to networks  whose interaction structure can take an arbitrary shape. More generally in this work we consider quantum complex networks,  mimicking real-world ones \cite{Nokkala2018}, as examples of complex environments \cite{Asenjo2017,Bello2019}. The role of the open system and of the network (environment) are played in the experiment by optical spectral modes interacting in a multimode non-linear process pumped by a femtosecond laser. \\
We show that the quantum optical platform, by using squeezing and entanglement correlations as resources along with continuous variables (CV) measurement \cite{Cai2017, Roslund2014}, is able to reproduce two crucial features of the system-environment dynamics. The first one is the energy exchange/dissipation, characterized by the spectral density of the environmental coupling \cite{Vasile2014}.  The second property is the quantum non-Markovianity (QNM) \cite{Rivas2010, Rivas2014,Breuer2009,Li2018,Laine2010}. Specifically we test the QNM as introduced by Breuer et al \cite{Breuer2009},  consiting of a  back-flow of information from the environment to the system  during the dynamics \cite{Breuer2016}. \\
While there exists number of experiments that have shown the ability to emulate open quantum systems and in particular to control  Markovian - non-Markovian transition  or dephasing effects \cite{Liu2011,Barreiro11,Cialdi17,Yu18,Liu18,Garcia-perez-2020}, these have been mainly implemented via single or few qubits. Only preliminary experimental studies \cite{Groblacher2015} of continuous variable (CV) systems have been reported and without any control of the open system dynamics. Here we take a significant step beyond these regimes by not only studying and controlling the dynamics of open system coupled to a multiparty CV environment but also shaping the fine-grained complex structure of the latter and finally experimentally testing the probing schemes for the environmental properties. Reservoir control and engineering are essential in quantum information tasks \cite{Harrington22, Menu22,Davidson22}  and the probed quantities, spectral density and quantum non-Markovianity, are the key features to look at  \cite{Li2019, Head-Marsden21,Gluza21,Shirai21,Paulson22b,Carrega22,Spaventa22}.


\section{Results}
\subsection{Mapping of the open quantum system into CV optical systems }
\begin{figure}[ht]
    \centering
    \includegraphics[width=11cm]{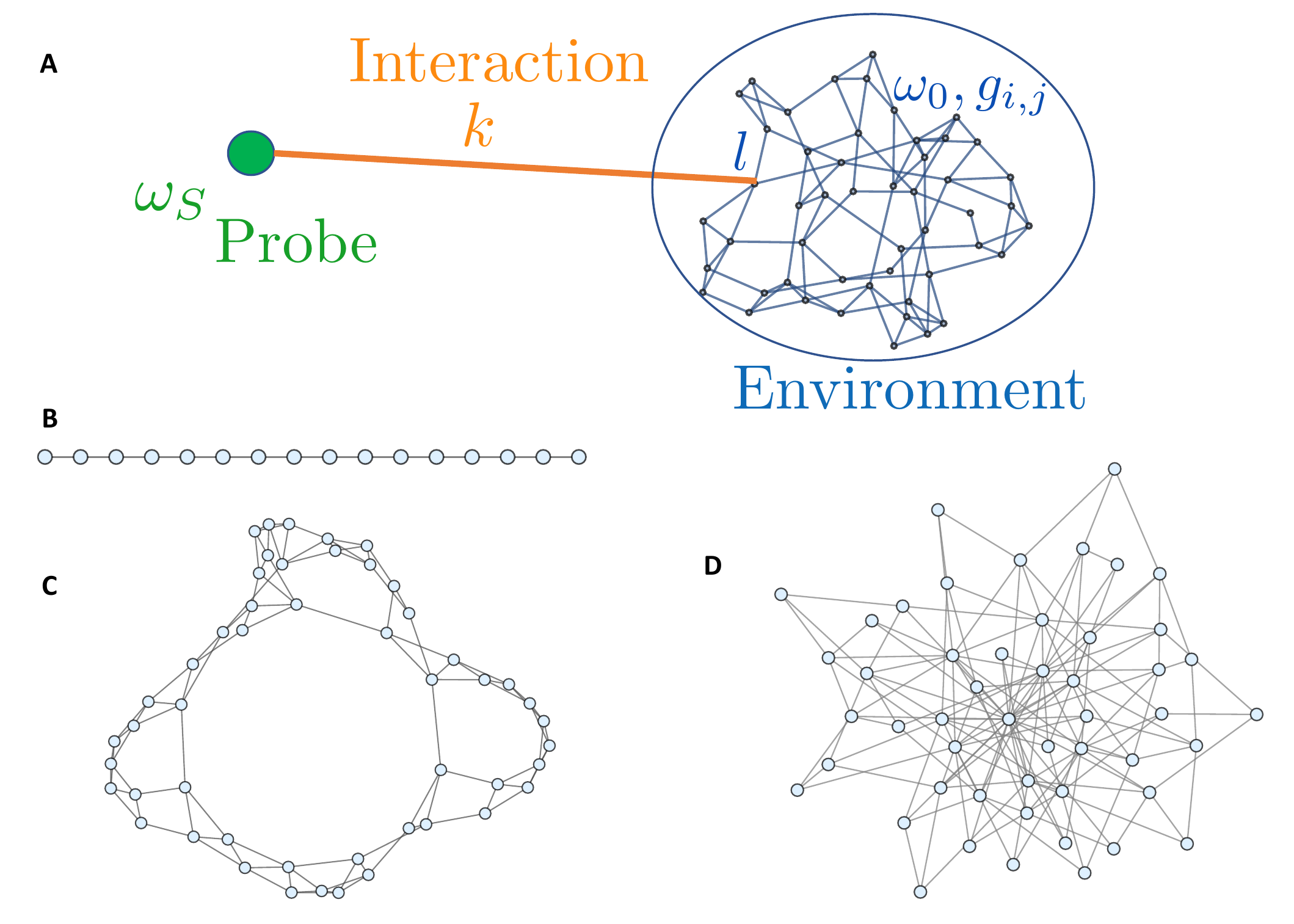}
    \caption{Structure of the global system and environments. \textbf{A}: Scheme of a global system where a probe (green) is attached to an environment network (blue) via a coupled strength $k$ (orange).
    \textbf{B-D}: Environment structures investigated in this work. \textbf{B}: linear structure used for the network 1, 2 and 3. These are networks of 16 nodes with periodic coupling strength such as network 1: $g_{i,i+1}=\{0.1; 0.05; 0.1; ...\}$; network 2: $g_{i,i+1}=\{0.1; 0.1; 0.05; 0.1; ...\}$; network 3: $g_{i,i+1}=\{0.1; 0.05; 0.025; 0.1; ... \}$. The coupling strengths between the probe and these three environments are the same: $k_{1-2-3}=0.01$.  \textbf{C} and \textbf{D} Structure that represents the environments called respectively network 4 and network 5 and both composed by $N=50$ nodes. They are two instances of the Watts-Strogatz (WS) and Barabasi-Albert (BA) model as defined in complex networks theory   \cite{Barabasi16,Newman18}. The coupling strengths are constant within  these two networks: $g_{i,i+1}=0.08$ for network 4 and $g_{i,i+1}=0.02$ for network 5 and we have $k_4=0.02$ and $k_5=0.004$. For each network, the nodes have the same frequency $\omega_0=0.25$. }
    \label{opandnetwork}
\end{figure}

The open quantum system we emulate (S, also named the probe) is coupled to an environment (E),  as shown in Fig. \ref{opandnetwork} A. S is a harmonic oscillator of frequency $\omega_s$. 
The environment is modelled by an ensemble of $N$ other harmonic oscillators coupled with each other via spring-like interactions. The coupling strength between oscillators $i$ and $j$ is denoted as $g_{ij}$. Without loss of generality, we consider the case where the frequencies of the oscillators in E are the same, $\omega_0$. We assume that S is coupled to only one node of the environmental network labelled $l$ (see Fig. \ref{opandnetwork}). The Hamiltonians of the network ($H_E$), of the system  ($H_S$) and of their interaction  ($H_I$) are \cite{Nokkala16,Nokkala2018}:
\begin{equation}
H_E=\bm{p}^T\bm{\Delta_\omega}\bm{p} + \bm{q}^T \sqrt{\bm{\Delta_\omega}^{-1}}\bm{A}\sqrt{\bm{\Delta_\omega}^{-1}}\bm{q} ; \quad H_S= \omega_S\left( \frac{p_S^2}{2} + \frac{ q_S^2}{2}\right) ; \quad H_I= k q_Sq_l
\label{HamiltonianTotal}
\end{equation}

where $\bm{A}$ is a symmetric and real matrix of size $(N)$ called the adjacency matrix of the environmental network, such that $A_{ij}=\delta_{ij}\omega_i^2/2-(1-\delta_{ij}g_{ij}/2)$,  $\bm{q}=( q_1,...q_N)^T$, $\bm{p}=(p_1,...p_N)^T$ and $\left(p_S,q_S\right)$ stands for renormalized quadrature operators\footnote{With $\left(\bm{q'},\bm{p'},q_S',p_S'\right)$ the usual quadrature operators of N+1 harmonic oscillators, they are defined as $\bm{q}^T=\bm{q'}^T\sqrt{\bm{\Delta_\omega}}$; $\bm{p}^T=\bm{p'}^T\sqrt{\bm{\Delta_\omega}^{-1}}$; $p_S=p_S'\sqrt{\omega_S^{-1}}$; $q_S=q_S'\sqrt{\omega_S}$.}.

Hamiltonians in Eq. (\ref{HamiltonianTotal}) are given with $\hbar=1$, and $m=1$. Also, in the rest of the article numerical values of couplings and frequencies are specified relative to a fixed (arbitrary) frequency unit. This is because on the one hand any possible value and unit can be chosen in the  simulations implemented via the optical system, and on the other hand the properties of the open system in presence of different environments are driven by the ratio between involved frequency and coupling terms $\{\omega_S,\omega_0,g,k \}$ rather than their absolute value.
We can easily derive the temporal dynamic of the full system, described by the  evolution matrix  $\mathscr{S}(t)$  such as:
\begin{equation}
    \bm{x}(t)=\mathscr{S}(t)\bm{x}(0).
\end{equation}
where $\bm{x}(t)=(q_s,\bm{q},p_s,\bm{p})$. More details about the derivation of  $\mathscr{S}$ can be found in  \cite{Nokkala2018}.
$H_E$,$H_S$ and $H_I$ being quadratic Hamiltonians leading to Gaussian processes, at each time $t$ the evolution matrix $\mathscr{S}(t)$ is symplectic. To study the energy transfer between the network and the system, some of the involved oscillators should be initialized in a non-zero energy state; such preparation can be included in the symplectic operation, $\mathscr{S}_{eff}(t)=\mathscr{S}(t)\mathscr{S}_{in}$,  so that the global process can be written as:  
\begin{equation}
\begin{split}
\bm{x}(t) & = \mathscr{S}_{eff}(t)\bm{x}_v(0)\\
 & = \bm{R}_1(t)\bm{\Delta}(t)\bm{R}_2(t)\bm{x}_v(0)=\bm{R}_1(t)\bm{\Delta}(t)\bm{x}_v(0).
\end{split}
\label{BM1}
\end{equation}
The second line is the Bloch-Messiah decomposition of the symplectic transformation where $\bm{R}_1$ and $\bm{R}_2$ are orthogonal matrices, corresponding to linear optics operations (or more generally to basis change), while $\bm{\Delta}$ is a diagonal matrix corresponding to squeezing operations \cite{Bloch1962}. As the initial state preparation is included in  $\mathscr{S}_{eff}$, $\bm{x}_v(0)$ are quadratures of a collection of oscillators in the vacuum states and $\bm{R}_2$ can be discarded.

We can obtain the same transformation of Eq. (\ref{BM1}) on quadratures of optical modes, using multimode squeezing ($\bm{\Delta}$) and mode basis change ($\bm{R}_1$). Experimentally, these  can be implemented in optical parametric process pumped by an optical frequency comb and measured via ultrafast shaped  homodyne detection \cite{Roslund2014,Cai2017,Nokkala2018}.
We can thus emulate the evolution of the open quantum system at a time $t$ by implementing the transformation $\bm{R}_1(t)\bm{\Delta}(t)$ on optical modes of different optical spectrum that play the role of the harmonic oscillators composing the network and the system. Starting from a bunch of initial modes to which the squeezing operation $\bm{\Delta}(t)$ is applied,  the linear optics operation $\bm{R}_1(t)$ corresponds to a basis change that can be realised by measuring the quadratures $\bm{x}(t)$ via the appropriate local oscillator shape in homodyne detection \cite{Cai2017,Nokkala2018}.
This platform can simulate the dynamic of an open quantum system in environments with tunable spectral features, induced by  environment. The environment can have any complex structure of correlations, as long as the number of optical modes  that can be detected in the mode-selective homodyne - i.e. the number of addressable harmonic oscillators- is large enough.
In the following we will show how our experimental system is able to simulate environments of different shape and size by recovering their specific features in the interaction with the open system. In particular we will recover the spectral density $J(\omega)$ \cite{Vasile2014}, that reveals the energy flow between the system and the environment, and the quantum non-Markovian behaviours \cite{Breuer2009}, i.e. an information back-flow from the environment to the system. 

In order to show the reconfigurability of our system we implement the networks shown in Fig. \ref{opandnetwork}. We set 3 different linear networks of 16 nodes with different periodic coupling strength \cite{Vasile2014}. Then we set networks of 50 nodes derived from complex-network models like the  Watts-Strogatz model,  characterized by short average path lengths and high clustering where we took rewiring probability $p_{WS}=0.1$ \cite{Watts1998}, and the Barabasi-Albert model characterized by a power law distribution of the degree with connection parameter set to $\kappa=2$ \cite{Albert02}.

\subsection{Spectral density for a given network}
\begin{figure}[ht]
    \centering
    \includegraphics[width=16cm]{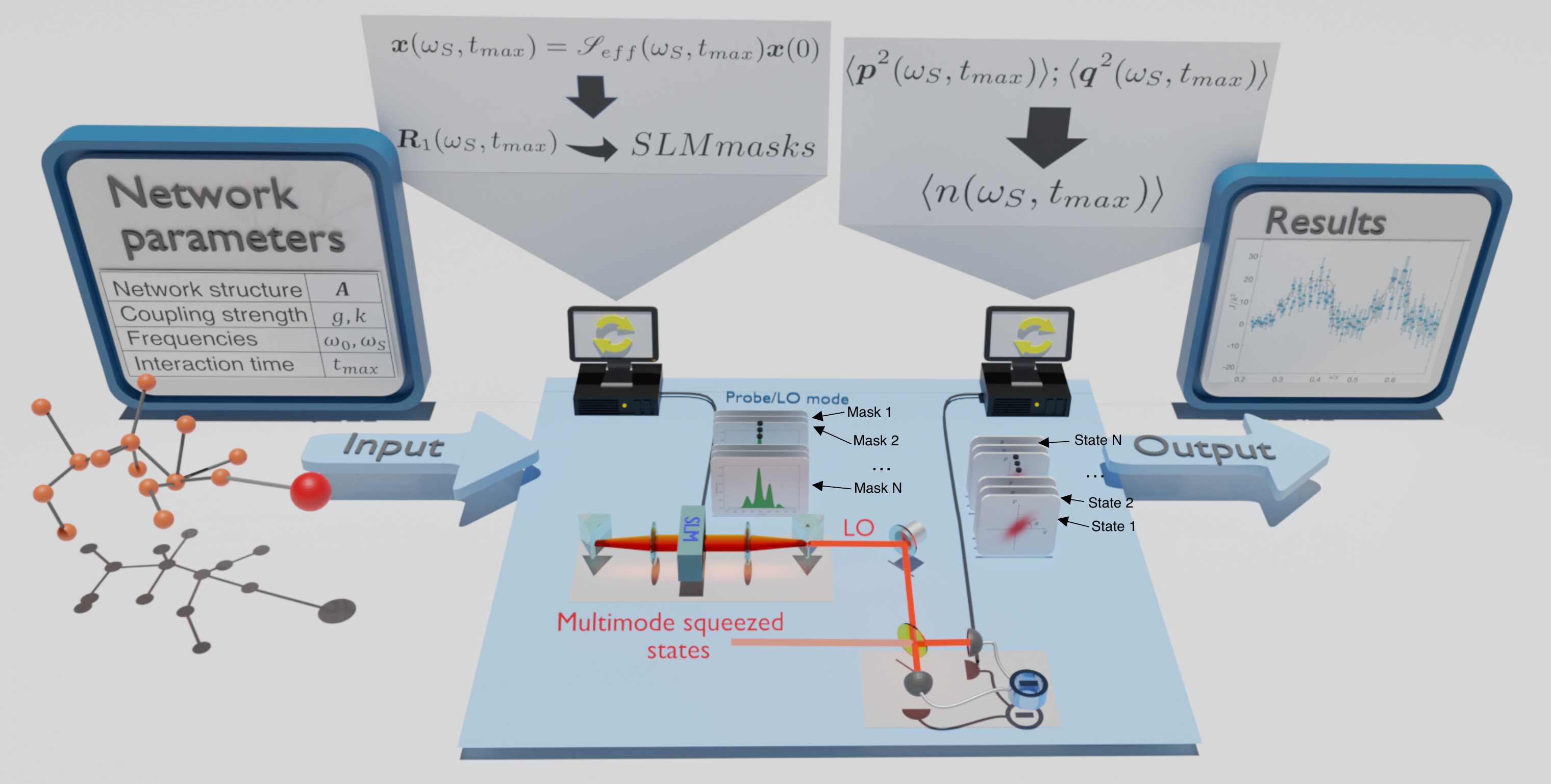}
    \caption{Protocol for the recovery of the spectral density function for a given environment. The parameters of a given network are used to calculate the quadrature dynamics of the system $S$ for different frequencies $\omega_{S}$ while the decomposition of such evolution gives the measurement setting, $\bm{R}_1(\omega_{S})$ is tied to a specific mask for the spatial light modulator (SLM) in the shaping of the local oscillator (LO). Measurement of $\langle \hat{q}_S^2 \rangle$ and $\langle \hat{p}_S^2  \rangle$ via the quadrature statistics are associated to the $\langle \hat{n}_S \rangle$ value and then to $J(\omega_{S})$.}
    \label{Protocol}
\end{figure}

\begin{figure}
    \centering
    \includegraphics[width=13cm]{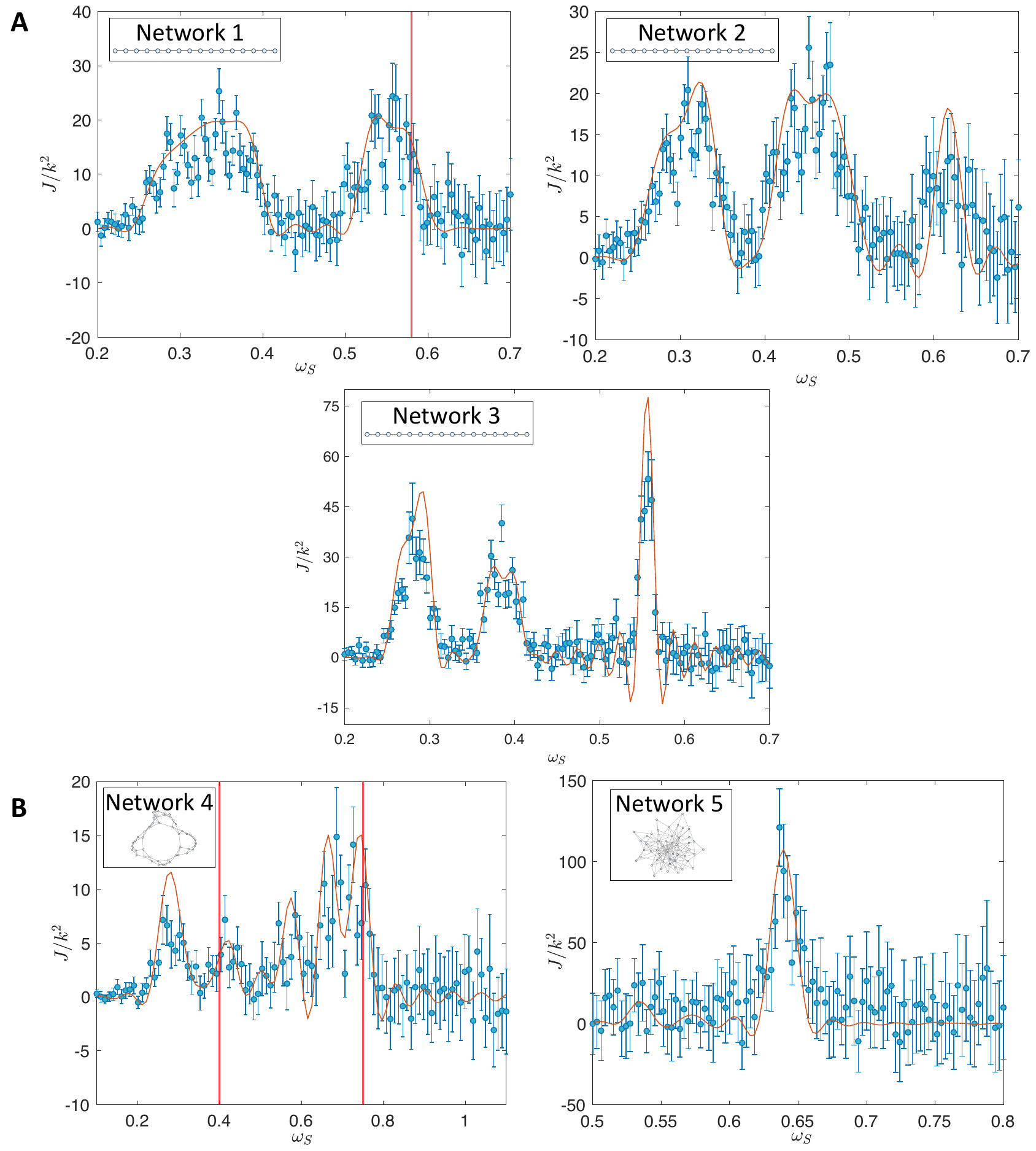}
    \caption{Spectral density measurements. Spectral density as a function of the frequency $\omega_s$ when the environment has a linear structure (\textbf{A}) or a complex structure (\textbf{B}). Orange lines: theoretical calculation; blue dots: experimental values recovered by probing the optical mode simulating the open system. Red vertical lines show the probe frequencies of network 1 and 4 for which QNM is investigated. 
    }
    \label{figureJ}
\end{figure}

The reduced dynamic of the system in presence of the environment can be derived by tracing out the environmental degrees of freedom from the evolution given by the total Hamiltonian $H_E+H_S+H_I$. In 
general the evolution of the open system takes the form of a non-unitary master equation for the state, or an equivalent generalized Langevin equation for the position observable, \cite{Gar04,Breu07,Weiss08,Vasile2014,Mascherpa20}   such as:
\begin{equation}
\ddot{q_S} + \tilde{\omega}_S^2 q_S + \int_{0}^{t} d\tau \gamma (t-\tau)\dot{q_S}=\xi(t),
\end{equation}
where $\tilde{\omega}_S$  is a renormalized system frequency and $\xi(t)$ is Langevin forcing of the system. The dissipation  and memory effect of the system are  featured by the damping kernel $\gamma (t)$. This can be equivalently characterized via the spectral density defined as:
\begin{equation}
    J(\omega_S)=\omega_S\int_0^{t_{max}}\mathrm{d}t\gamma(t) \cos(\omega_S t)
    \label{Janalytical}.
\end{equation}
The value of the spectral density at a given frequency $\omega_S$ shows the strength of the energy flow between the system and the environment, e.g. its damping rate. Specific network structures of the environment are characterized by different shapes of $J(\omega_S)$ that can be easily recovered by calculating the evolution of the system plus the environment  via the total Hamiltonian in  Eq. (\ref{HamiltonianTotal}) and getting $\gamma(t)$ from the reduced dynamics of the system. 
It should be noted that $ J(\omega_S)$ in Eq. (\ref{Janalytical}) is normally defined with $t_{max}=\infty$ but as here we are dealing with finite environments we set a finite $t_{max}$, which can be considered as the time the open system interacts with all the elements of the networks before the revival dynamics arises due to the finite size effects of the environment (see Methods).  \\
 For unknown network structures, the shape of  $J(\omega_S)$ can be recovered by probing the excitation number of the system $S$ that interacts with the environment \cite{Vasile2014,Nokkala16,Nokkala2018} at different frequencies $\omega_S$ (see Eq. (\ref{probJnx}) in Methods). This is the approach we follow in the experimental simulation: given all the network parameters, the experimental setup composed by different optical modes can implement the quadrature  evolution $\bm{x}(t)$ of the network plus the system   in  Eq. (\ref{BM1}). We get the value of $J(\omega_S)$ by monitoring the excitation number $\langle \hat{n_S} \rangle$, that can be  recovered from homodyne measurements of $\langle \hat{q_S}^2 \rangle$ and $\langle \hat{p_S}^2  \rangle$ and we compare the results with the expected theoretical shape. The protocol is shown in Fig.  \ref{Protocol}.  The experimental data in Fig. \ref{figureJ}  are obtained from homodyne measurements of the mode corresponding to (S) having interacted with (E) until $t_{max}$. 
Each dot is the value of $J(\omega_S)$ recovered when the environment interacts with the system at a specific frequency $\omega_S$. This corresponds to a given measurement setting, i.e. to a given basis change $\bm{R}_1(t_{max},\omega_S)$ in Eq. (\ref{BM1}),  and in particular to a given Local Oscillator (LO) spectrum in the homodyne measurement, set by the SLM (Spatial Light Modulator) mask. 
 One dot on each curve is the average of  20 measurements and the error bars are obtained from the standard deviation. The theoretical curves are calculated from Eq. (\ref{Janalytical}) given the parameters of the networks shown in Fig. \ref{opandnetwork} (see Methods). Except for some noise due to the instabilities of the experimental system, the experimental data shown in Fig. \ref{figureJ} match the shapes of the theoretical curves for all complex environments.


\subsection{Quantum non-Markovianity}
In addition we simulate quantum Markovian or non-Markovian behaviour of the studied networks for some parameters range. In this work we use the  definition of quantum non Marknovianity (QNM) introduced by Breuer et al in \cite{Breuer2009}, where the memory effect in the system is associated to a back flow of information from E to S. A Markovian process continuously tends to reduce the distinguishability between any two quantum  states of a given system, in  a non-Markovian behavior it tends to increase, so that the flow of information about such distinguishability is reversed from the environment to the system. 
The original definition of QNM based on trace distance between the states can be expressed,  in the case of Gaussian states, via the fidelity as proposed in \cite{Breuer2009,Vasile2011}. Thereforewe use the following  QNM witness:
\begin{equation}
    \mathcal{N}=\max_{\rho_1,\rho_2}\int_{\frac{\partial F}{\partial t}<0}\mathrm{d}t\Big(-\frac{\partial F}{\partial t}\Big),
    \label{NM Witness}
\end{equation}
where $(\rho_1,\rho_2)$ is a pair of states of S and $F$ their fidelity. 
Experimentally only a finite set of quantum states can be accessed, we have then chosen the two experimentally accessible states $\rho_1,\rho_2$ that minimize the fidelity at $t=0$, in order to have the maximal sensitivity in the QNM witness $ \mathcal{N}$. Such states are two vacuum squeezed states that are squeezed along two orthogonal directions (see Methods). 
\\
The quantum non-Markovianity can be associated to specific structures of the spectral density \cite{Vasile2014}, in particular it has been shown that maximal values of $ \mathcal{N}$  are reached at the edges of a band-gap in spectral density, where the band-gap is a region where  $J(\omega_S)$ is close to zero. 
In this protocol we focus on  some given values of the probe frequency where the spectral density of specific networks (linear and WS) have particular features: large values, or being within or at the edge of the gap.   
For each time $t$ the protocol is performed for two different input states $\rho_{1sq}$ and $\rho_{2sq\perp}$, in order to measure their fidelity.  The first state  $\rho_{1sq}$ has squeezing of $-1.8$ dB along the $q$ quadrature and antisqueezing of $+2.9$ dB, and the second $\rho_{2sq\perp}$ has squeezing of $-1.3$dB along the $p$ quadrature and antisqueezing of $+2.4$ dB. The fidelity between the two states is calculated via the covariance matrix of the two recovered via the measurement of  $\langle \hat{q_S}^2 \rangle$ and $\langle \hat{p_S}^2  \rangle$.

\begin{figure}[ht]
    \centering
    \includegraphics[width=13cm]{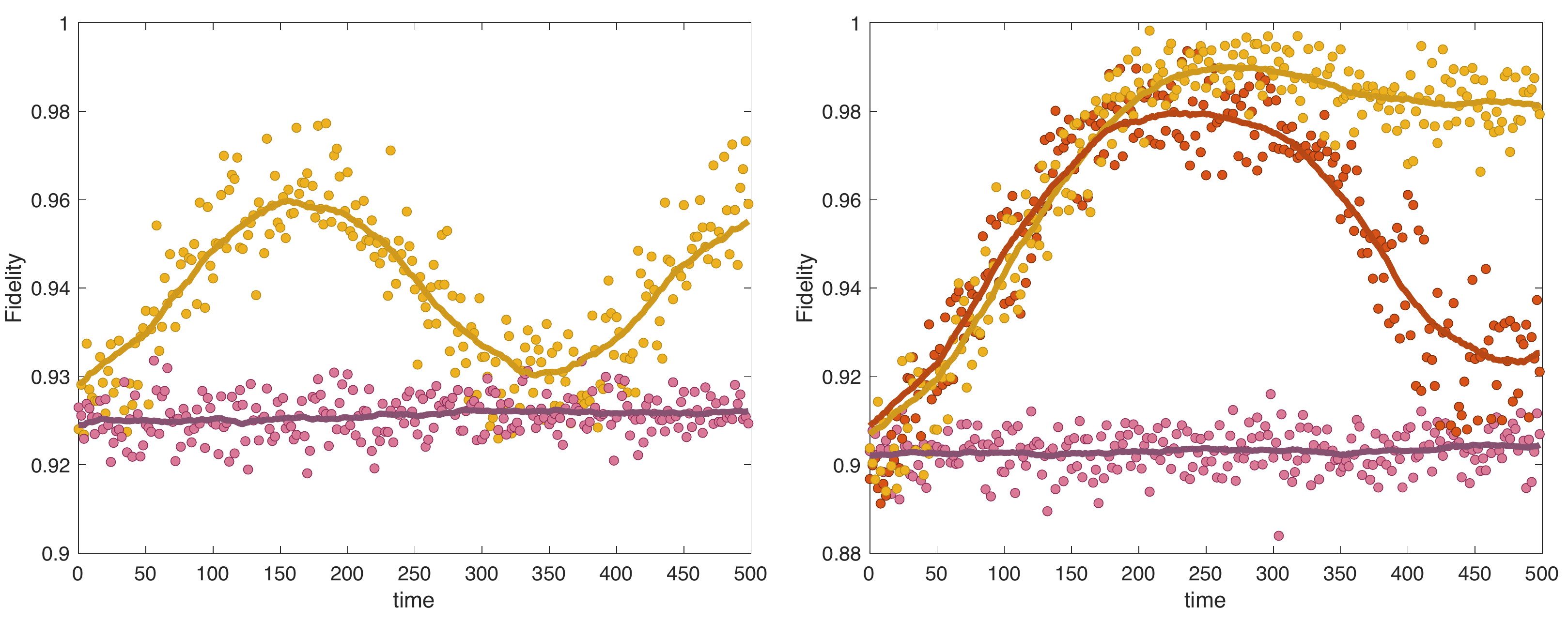}
    \caption{ Fidelity as a function of the time parameter for several probe frequencies. (\textit{left}) The environment takes the form of network 1 and pink and yellow dots are respectively for $\omega_S=0.7$ and $\omega_S=0.58$. (\textit{right}) The environment takes the form of the Watts-Strogatz network,  pink dots are measured for $\omega_S=0.9$, red dots for  $\omega_S=0.75$ and the yellow dots for  $\omega_S=0.4$. Solid lines are obtained via averaging over 50 points. }
    \label{FigureF}
\end{figure}

Fig. \ref{FigureF} shows the fidelity measurement as a function of time $t$ for the linear periodic network called network 1 and the Watts-Strogatz network at 
given values of frequency $\omega_S$. For both environments we can observe a back flow of the information in the system for some of the  monitored frequencies of the probe. When the environment takes the form of network 1, the system is non-Markovian for $\omega_S=0.58$, at the edge of the gap \cite{Vasile2014}, while no information exchange is perceived between the system and the environment for $\omega_S=0.70$. At such frequency, the dynamics of the system can be interpreted as unitary,  as shown by a value of $J(\omega_S)$ close to $0$ in Fig. \ref{figureJ} A. In the case of the Watts-Strogatz network, non-Markovianity is observed for $\omega_S=0.4$ and $\omega_S=0.75$ with a larger information back-flow for the latter, while no information exchange is noticeable at $\omega_S=0.9$. In order not to overestimate the witness value because of high frequency fluctuations  in the experimental data, the derivative of Eq. \ref{NM Witness} is evaluated on averaged curves ( solid lines in Fig. \ref{FigureF} ). The obtained values of $\mathcal{N}$ are gathered in the Table \ref{tableF}.

\begin{table}[ht]
    \centering
    \begin{tabular}{ |c|c|c|c| } 
    \hline
    $\omega_S$ &  $\mathcal{N}$ of network 1 &  $\mathcal{N}$ of  WS network\\
    \hline
    0.4 & - & 0.013 \\ 
    0.58 & 0.041 & - \\ 
    0.7 & 0.012 & - \\
    0.75 & - & 0.06 \\
    0.9 & - & 0.012 \\
    \hline
    \end{tabular}
    
    \caption{Non Markovian witnesses for network 1 and 4.}
    \label{tableF}
\end{table}



\section{Discussion}
In summary we have experimentally demonstrated the simulation of an open system coupled to complex network environments of different shape. Any network shape can be engineered and probed in the actual platform. In particular we have shown probing techniques for the spectral density of the environmental coupling and  of the quantum non-Markovianity.
Our platform is the first experimental setup where continuous variable open systems with engineered environment are tested. It goes beyond the few-qubits implementation by controlling a multipartite system with up to 50 components. The environment and system size can be increased in future experiments by considering both spectral and time multiplexing \cite{Kouadou2022}, moreover non-Gaussian interaction can be added \cite{Ra20,Wal2020}.
Applications are relevant in the context of quantum information technologies.  Dissipation phenomena in energy transfer and in particular vibronic dynamic can be mapped via the demonstrated experimental apparatus \cite{Mattioni21,Caycedo2022, Nussuler22}  opening the way to the test of artificial light-harvesting architecture. Moreover we can test engineered environment to enhance quantum thermal machines \cite{Manzano19,Kloc21}. Finally, we can explore different probing schemes like multiple harmonic oscillators (measured modes) coupled with different partitioning of the environment, via weak or strong interaction.  We can then test paradigmatic collective phenomena, like quantum phase transitions and quantum synchronization \cite{Minganti2018, Manzano13,Cabot19} and the emergence of classical world from the quantum one as the  effect of the interaction with a structured environment \cite{Galve16,Le19,Foti19}.


\section{Methods}

\subsection{The experimental setup}

\begin{figure}[ht]
    \centering
    \includegraphics[width=13cm]{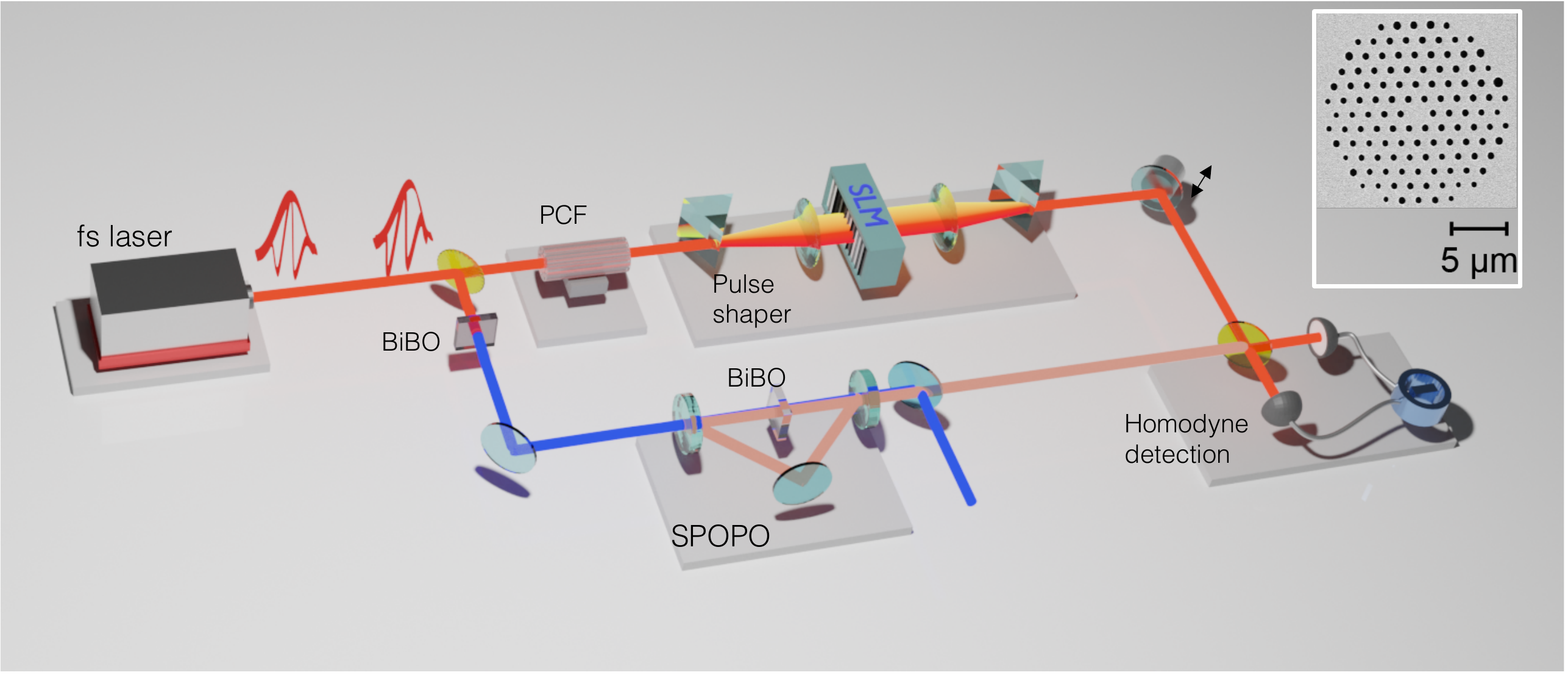}
    \caption{Scheme of the experimental setup. The beam from the femto-second laser source is divided in two paths: the first one to produce multimode squeezed light \{second-harmonic generation + SPOPO cavity\} and the second one to spectrally shape the local oscillator \{PCF + pulse shaper\}. The two beams are then recombined for the homodyne detection. Inset: electron micrograph of the end face of the PCF.}
    \label{FigureSPOPO}
\end{figure}

A general scheme of the experiment  is shown in Fig. (\ref{FigureSPOPO}). The main goal is to produce highly multimode non classical light. Frequency doubled pulses from a Ti:Sapphire laser are propagated in an OPO cavity with round trip time matched to the pump pulse train cycle time. The non linear crystal inside the cavity is a BiBO crystal of length $0.2$ mm. The repetition rate of the pulse train is $75$MHz and pulses have a duration of about $100$ fs. Correlations appear among many spectral modes of the frequency
comb of the down converted light, giving rise to a squeezed vacuum state at $795$ nm central wavelength with a highly multimode structure \cite{Roslund2014,Cai2017}. This multimode squeezing structure can be measured via a spectrally resolved homodyne detection. A pulse shaper, located upstream, enables to set the spectral mode of the local oscillator in the homodyne detection, which is then mode probed in the detection.  We find that the squeezed modes spectra are described by a set of Hermite-Gaussian functions. The first one of the series is a Gaussian function with FWHM of around 6.5 nm. 
 The quadratures evolution of the involved optical modes in the parametric process is then described by an equation of the same form of the Eq. (\ref{BM1}). The harmonic oscillators initially in a vacuum state and whose quadratures are squeezed as $\bm{\Delta}(t)\bm{x}(0)$ are, in the actual setup, the squeezed modes.

The  quadratures $\bm{x}(t) =\bm{R}_1(t)\bm{\Delta}(t)\bm{x}(0)$ are the ones accessed via homodyne detection. The transformation $\bm{R}_1(t)$ describes the basis change from the modes with Hermite-Gauss spectral shapes to the measured modes \cite{Roslund2014,Cai2017,Ferrini15}. 
In order to exactly match the evolution of the multimode optical system with the one of the network plus the probe, we have to set  both the squeezing values in $\bm{\Delta}(t)$ and the detected pulse shapes given by $\bm{R}_1(t)$. In the present experiment, numerical analysis showed that the probing of $ J(\omega_S)$ and of $\mathcal{N}$ are not very sensitive to changes in the squeezing values in $\bm{\Delta}(t)$. In these protocols the numbers and values of non-zero  diagonal terms in $\bm{\Delta}$ set the 
number of oscillators that are initially in a not-vacuum state and the value of their excitation numbers.
Interactions between the oscillators and their evolution can be established  via $\bm{R}_1(t)$. Thus only $\bm{R}_1(t)$ is modified accordingly to the dynamics we have to simulate.
In both protocols (probing of spectral density and quantum non-Markovianity) we follow the spread of energy/information from few initially populated harmonic oscillators to a large number of harmonic oscillators in the networks.

So if the number of the harmonic oscillators that can be initially set in a not-vacuum state is limited by the number of produced squeezed modes, the number of harmonic oscillators that can be reached by some excitation (i.e. the number of the total oscillators in the networks) is limited by the number of spectral modes we can measure, as this is what limit the size of the matrix $\bm{R}_1(t)$.
A larger number of spectral modes to measure means a larger spectrum to be shaped for the local oscillator field.
So in the end the number of the simulated harmonic oscillators depends on the capability  of the pulse shaper which is here limited by the optical complexity \cite{Monmayrant2010} and the spectral width of the field that is used as local oscillator. In order to not be limited by the latter, the local oscillator, that is derived from  the main laser source  is  broadened with a 2~cm-long all-normal dispersion photonic crystal fiber (PCF), before entering the pulse shaper stage. In such a fiber, the broadening mechanism only relies on self-phase modulation, which is known to be low-noise \cite{Abdolghader2020}.
If needed in future experiments and protocols, $\bm{\Delta}(t)$ can be opportunely controlled via the shaping of the pump in the parametric process \cite{Arzani18}.   
\\

\subsection{Protocol}
\subsubsection{Spectral density measurement}
The protocol is based on the bosonic resonator network mapping established by J. Nokkala et al \cite{Nokkala2018}. In table \ref{recap} are gathered the  equivalent items involved in the mapping to emulate network of bosonic oscillator.
\begin{table}[!ht]
    \center
    \begin{tabular}[b]{|c|c|c|}
    \hline
    Network component & Quantum network & Experimental implementation \\
    \hline \hline
    Node& Quantum harmonic oscillator & Optical mode  \\
    Link & Coupling strength & Entanglement/basis change  \\
    Addressing a node & Local measurement & Pulse shaping and projective measurement\\
    \hline
    \end{tabular}
    \caption{Mapping of experimental implementation for the quantum network for open quantum system.}
    \label{recap}
\end{table}

As shown in Fig. \ref{Protocol}, the protocol is carried out in two stages:  first, basis change and accordingly the mask characteristics setting the probe measurement are computed by a Mathematica code and second, the variances $\langle \hat{q_S}^2 \rangle$ and $\langle \hat{p_S}^2  \rangle$ are measured. The process is as follows:
\begin{itemize}
    \item Interaction time $t_{max}$, environment structure, coupling strength $g$ and $k$ and the frequency $\omega_0$ are set and entered in the code.  At first, the damping kernel $\gamma(t)$ is numerically calculated and then we selected as $t_{max}$ a value where the resulting solution is flat and close to 0. The $\gamma$ computation for the considered environments are in the supplementary material.\\
    For sufficiently short times the system cannot resolve the different frequencies of the network, making the spectral density a continuous function of frequency in this regime, as seen in Fig.~\ref{figureJ}. For some networks the spectral density additionally assumes a constant form for a transient where the shape is not sensitive to small differences in interaction time. The set interaction time for 
    each network are : $t_{max,1}=t_{max,2}=t_{max,3}=150$, $t_{max,4}=90$ and $t_{max,5}=250$.
    \item   A set of matrices  $\mathscr{S}_{eff}(t_{max},\omega_s)$ is  evaluated  for 120 values of $\omega_s$ in the range $\{0.2,0.7\}$ for network 1, 2 and 3. Then 100 matrices are also computed in the ranges $\{0.1,1.1\}$ and $\{0.5,0.8\}$ for respectively the network 4 and 5. The matrices $\bm{R}_1(t_{max},\omega_s)$ and $\bm{\Delta}(t_{max},\omega_s)$ are obtained from the Bloch Messiah decomposition.
    \item From $\bm{R}_1(t_{max},\omega_s) $ we can derive the spectral mode corresponding to the system/probe of frequency $\omega_s$ having interacted with the network for the time $t_{max}$. The corresponding optical spectrum of the Local Oscillator is   shaped via the  SLM masks. The average values $\langle \hat{q_S}^2 \rangle$ and $\langle \hat{p_S}^2  \rangle$ are obtained via homodyne detection. 
    \item  The average photon number $\langle n_S(t_{max})  \rangle$ of the system is derived so that we can get $J( \omega_s)$ from the following equation \cite{Nokkala2018}
     \begin{equation}\label{probJnx}
         J( \omega_s)=\dfrac{\omega_s}{t_{max}}\ln \Big(\dfrac{N(\omega_s)-\langle n_S(0)  \rangle}{N(\omega_s)-\langle n_S(t_{max}) \rangle} \Big),
     \end{equation}
    where $N(\omega_s)=\left(\mathrm{e}^{\omega_S/T}-1\right)^{-1}$ is the thermal average boson number with $T$ being the temperature of the environment.
\end{itemize}

\subsubsection{Quantum non-Markovianity}

Although the emulated total system dynamic remains unchanged, the way to highlight QNM is slightly different than the way to recover the spectral density function. 

\begin{itemize}
\item Environment structure, coupling strength $g$ and $k$, the frequency $\omega_0$ and the probe frequency $\omega_s$ are set and entered in the code. 
\item   A set of matrices  $\mathscr{S}_{eff}(t,\omega_s)$ is  evaluated  for 251 values of $t$ in the range $\{0,500\}$. The same set is applied to two different input states for the probe/system oscillator,  $\rho_{1sq}$ and $\rho_{2sq\perp}$ consisting in two vacuum states squeezed along two orthogonal directions. The two are naturally encoded in the first two modes $HG_{0}$,$HG_{1}$ of the Hermite-Gauss series that diagonalize the parametric down conversion Hamiltonian, $\rho_{HG_{0}}(0)=\rho_{1sq}$;  $\rho_{HG_{1}}(0)=\rho_{2sq\perp}$.
\item The set of SLM masks corresponding to the temporal evolution of the two initially squeezed oscillators are evaluated.
\item Homodyne measurements are used for the evaluation of the states $\rho_{HG_{0}}(t)$ and $\rho_{HG_{1}}(t)$  at time t and their fidelity.
\end{itemize}


\section*{Acknowledgements}

\section*{Funding Statement}

This   work   was   supported   by   the   European   Research Council under the Consolidator Grant COQCOoN (Grant No.  820079).
S.M. acknowledge financial support from the Academy of Finland via the Centre of Excellence program (Project No. 336810 and Project No. 336814). J.N. acknowledges financial support from the Turku Collegium for Science, Medicine and Technology as well as the Academy of Finland under project no. 348854. R.Z. acknowleges funding from  the Spanish State Research Agency, through the María de Maeztu project CEX2021-001164-M and the QUARESC project PID2019-109094GB-C21 (AEI /10.13039/501100011033), and CAIB QUAREC project (PRD2018/47).

\section*{Competing interests} 
The authors declare that they have no competing interests.

\bibliography{bib}
\newpage
\section*{Supplementary Files}
\subsection{Witnessing non-Markovianity with pure squeezed states}

The used witness $\mathcal{N}$ of quantum non-Markovianity (QNM), specified in Sec.~$(2.3)$ of the main manuscript, involves a maximization over pairs of distinct initial states. Experimental quantification will naturally limit the possible pairs to experimentally accessible states, such as pure squeezed states characterized by the magnitude $r_i$ and phase $\varphi_i$ of squeezing where the index $i\in\{1,2\}$ indicates the two states.

Here plausibility is provided for the following claims: non-Markovianity is witnessed independently of the initial phase difference $\varphi_0:=\varphi_1-\varphi_2$; squeezing opposite quadratures, as we have done, can be expected to give the highest value for $\mathcal{N}$ in this set of states. In particular, if the open system dynamics is found to be (non-)Markovian for one choice of $\varphi_0$ it will be found (non-)Markovian for all values $\varphi_0\neq 0$; only the numerical value of $\mathcal{N}$ will change.

The key argument is that the choice of $\varphi_0$ maximizing distinguishability at $t=0$ maximizes it also for any time $t>0$ whereas the qualitative behavior is independent of $\varphi_0$. As will be seen, this is a consequence of the following points, suggested both by intuition and numerical simulations:
\begin{enumerate}
    \item the phase difference remains nearly constant: $\varphi(t)\approx\varphi_0$;
    \item $r_i(t)$ is nearly independent of $\varphi_i$;
    \item qualitative behavior of $r_i(t)$ is not sensitive to $r_i(0)$.
\end{enumerate}

We start from the fidelity $F$ between the two states. It reads
\begin{equation}
F=\frac{2}{\sqrt{2(1+\cosh{2r_1}\cosh{2r_2}-\cos{\varphi_0}\sinh{2r_1}\sinh{2r_2})}}.
\end{equation}

As anticipated only the relative phase matters. We turn our attention to $F(t)$. Because of point $1.$ we may substitute $\varphi(t)$ with $\varphi_0$. Then the squeezing parameters are the only source of non-monotonicity.
\begin{equation}
F(t)\approx\frac{2}{\sqrt{2(1+\cosh{2r_1(t)}\cosh{2r_2(t)}-\cos{\varphi_0}\sinh{2r_1(t)}\sinh{2r_2(t)})}}.
\end{equation}

We consider two limiting cases: identical phase $\varphi_0=0$ and opposite phase $\varphi_0=\pi$. Because of point $2.$ the resulting expressions can be directly compared. We get for the two, respectively,
\begin{equation}
F(t)\approx\frac{2}{\sqrt{\cosh^2(r_1(t)-r_2(t))}},\quad F(t)\approx\frac{2}{\sqrt{\cosh^2(r_1(t)+r_2(t))}}.
\end{equation}
As expected, these are the maximal and minimal values of fidelity, respectively, in the interval $\varphi_0\in[0,\pi]$ because there $F(t)$ is monotonically decreasing. This follows from the non-negativity of the hyperbolic functions and continuity in $\cos{\varphi_0}$ which is monotonous in $[0,\pi]$. Given point $3.$, full contribution of non-monotonicity from both squeezings is achieved only in the latter case and therefore $\mathcal{N}$ is maximized when $\varphi_0=\pi$.

In the special case $r_1=r_2:=r$ point $3.$ becomes unnecessary. Here fidelity in general and in the case $\varphi_0=\pi$, respectively, simplifies to
\begin{equation}
F(t)\approx\frac{2}{\sqrt{\cosh{(4r(t))}(1-\cos{\varphi_0})+\cos{\varphi_0}+3}},\quad F(t)\approx\mathrm{sech}2r(t).
\end{equation}

There are caveats. First of all this is clearly not a proof but rather justifies why our choice for the initial states and the interpretation of the results are reasonable. Second, the equations above implicitly assume pure states. Since we have used a weak coupling and pure state for the network purity can be expected to remain high for the considered times and fidelity should behave as above. This has been checked with numerical simulations.

\begin{figure}[ht!]
    \centering
    \includegraphics[width=15cm]{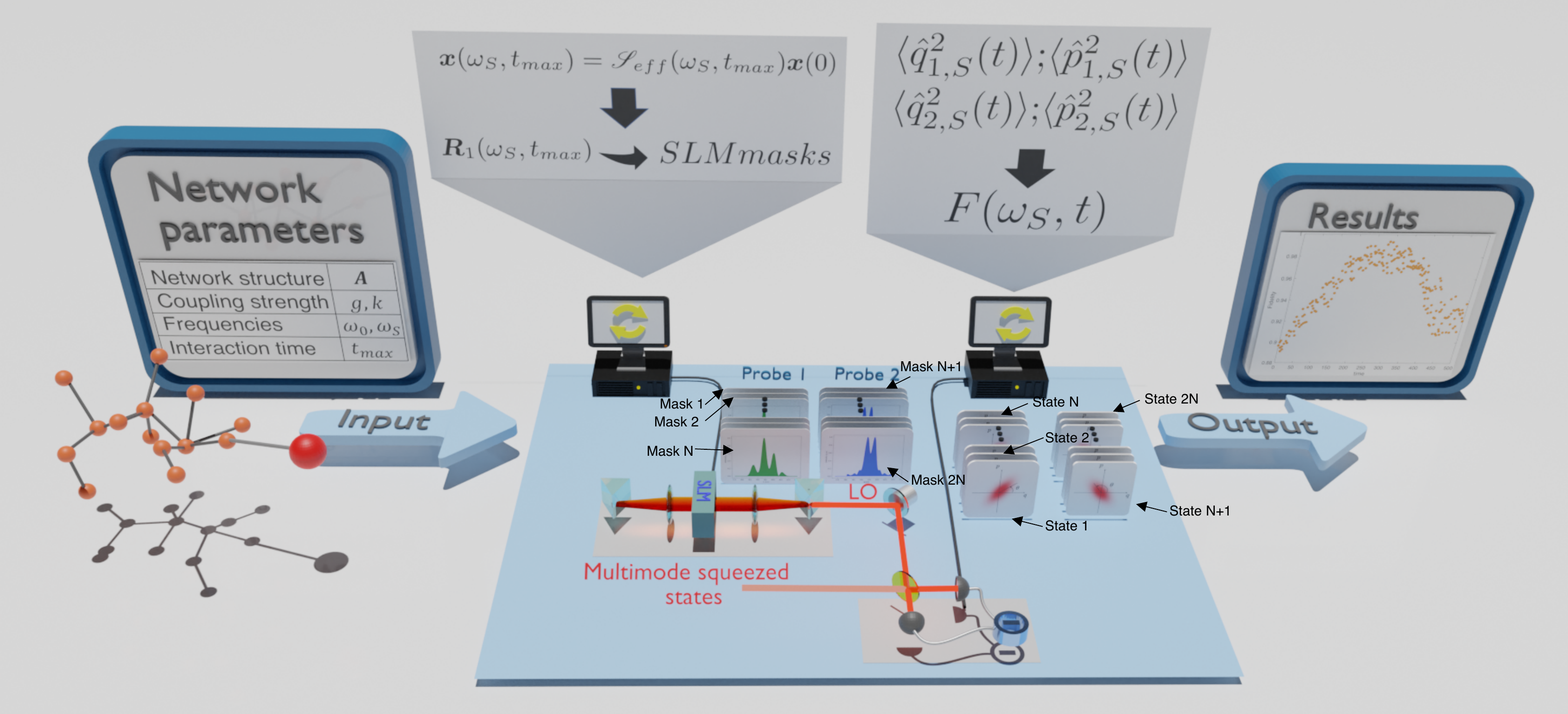}
    \caption{Protocol for the determination of the fidelity and consequently the BLP witness as a function of time. Here, two types of computed spectrum are sent to the SLM: one related to state $\rho_{1sq}$ (green) and another one related to state $\rho_{2sq\perp}$ (blue). The tomography of state, related to a given time $t$, is performed by the homodyne detection and is then determined the corresponding fidelity.}
    \label{supp1}
\end{figure}

\newpage

\begin{figure}
    \centering
    \includegraphics[width=15cm]{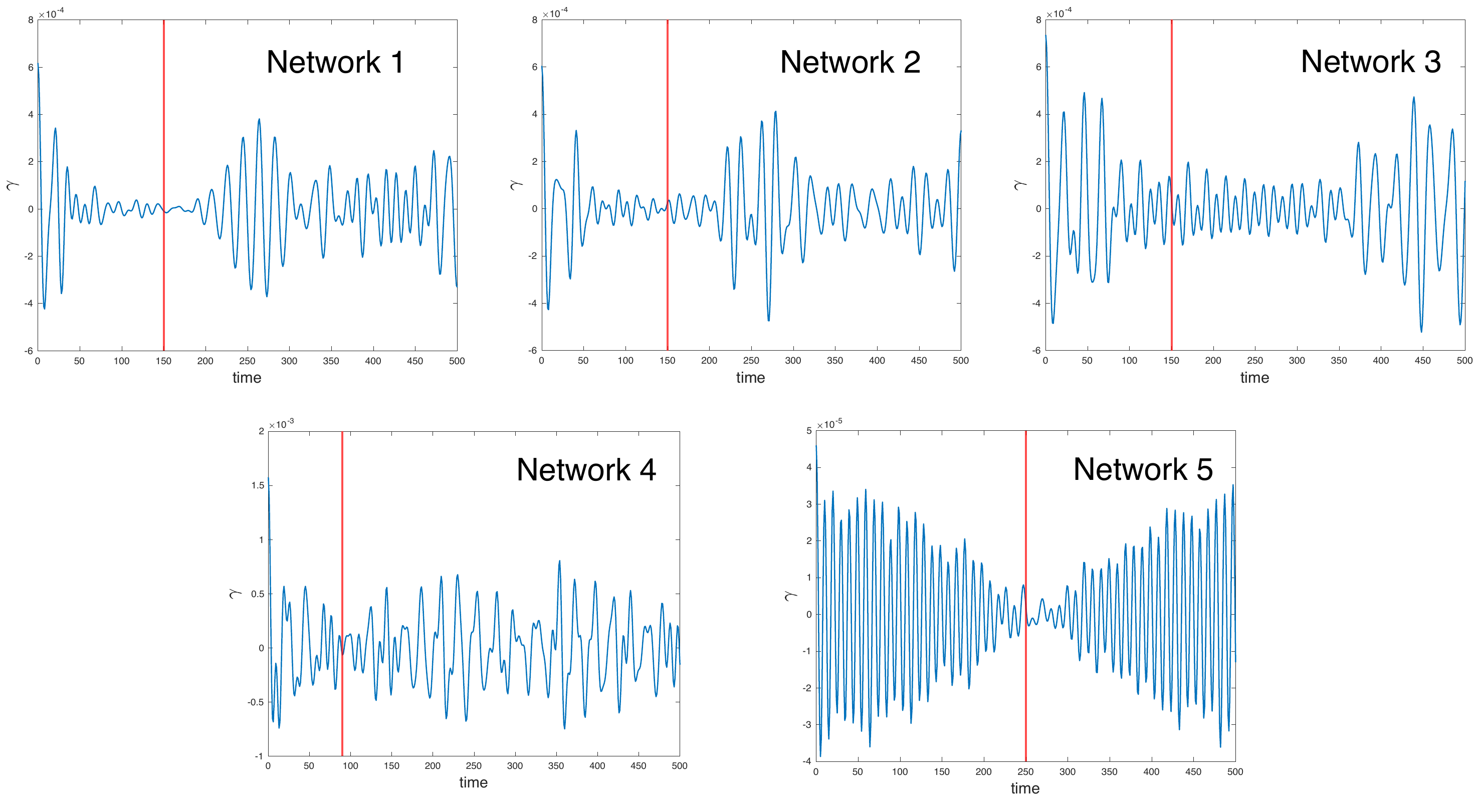}
    \caption{Analytical simulation of damping kernel as a function of time for all networks. For each network implemented as environment of the probe, the resulting damping kernel exhibits the same shape: first, it is observed a decay of the oscillations amplitude until reaching then a more or less long plateau where is set $t_{max}$. Let's notice that for the linear networks the plateau is more easily distinguishable than for the complex network environment. The red lines show the position of $t_{max}$ selected for the simulation of the different total system dynamics.}
    \label{supp2}
\end{figure}

\end{document}